\documentclass{article}

\usepackage[T1]{fontenc} 
\usepackage[utf8]{inputenc} 
\usepackage{ismir,amsmath,cite,url}
\usepackage{graphicx}

\usepackage{color}

\usepackage[firstpage]{draftwatermark}
\definecolor{lightgray}{rgb}{0.9,0.9,0.9}
\definecolor{darkgray}{rgb}{0.4,0.4,0.4}
\SetWatermarkFontSize{12pt}
\SetWatermarkScale{1.1}
\SetWatermarkAngle{90}
\SetWatermarkHorCenter{202mm}
\SetWatermarkVerCenter{170mm}
\SetWatermarkColor{darkgray}
\SetWatermarkText{Late-Breaking / Demo Session Extended Abstract, ISMIR 2022 Conference}

\def\lbd{}	

\title{Augmented Reality Visualization for Musical Instrument Learning}

\twoauthors
{Frank Heyen} {VISUS, University of Stuttgart \\ frank.heyen@visus.uni-stuttgart.de}
{Michael Sedlmair} {VISUS, University of Stuttgart \\ michael.sedlmair@visus.uni-stuttgart.de}

\def\authorname{F. Heyen and M. Sedlmair}

\usepackage[bookmarks=false,pdfauthor={\authorname},pdfsubject={\papersubject},hidelinks]{hyperref}

\begin{document}

\maketitle
\begin{abstract}
We contribute two design studies for augmented reality visualizations that support learning musical instruments.
First, we designed simple, glanceable encodings for drum kits, which we display through a projector.
As second instrument, we chose guitar and designed visualizations to be displayed either on a screen as an augmented mirror or as an optical see-through AR headset.
These modalities allow us to also show information around the instrument and in 3D.
We evaluated our prototypes through case studies and our results demonstrate the general effectivity and revealed design-related and technical limitations.
\end{abstract}
\section{Introduction}

\begin{figure*}[t]
 \centerline{
 \includegraphics[width=\linewidth]{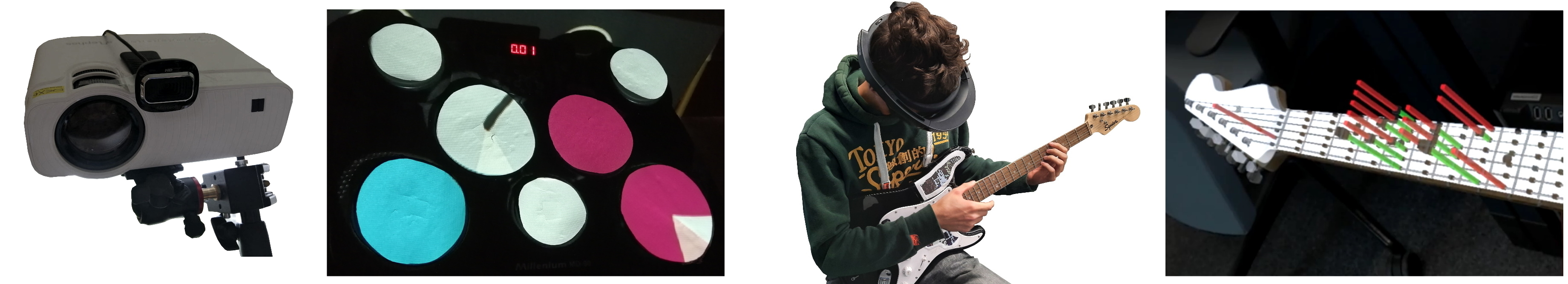}}
 \caption{
    The two augmented reality setups we explored: 
    Left: Using a projector on a tripod, we display visualizations on the pads of a drum kit.
    Right: A user wearing an optical see-through AR headset sees visualizations on a guitar's fretboard.
 }
 \label{fig:teaser}
\end{figure*}

Learning a musical instrument requires time and effort, even more without appropriate feedback. 
While some learners are taught by professional teachers, many are self-taught for at least some period of time, and during practice between lessons no professional feedback is available.
Several online services and apps\footnote{\label{apps}For example \href{https://rocksmith.ubisoft.com/}{rocksmith.ubisoft.com}, \href{https://yousician.com/}{yousician.com}, and\\ \href{https://synthesiagame.com/}{synthesiagame.com}} try to fill this gap, but their feedback is often highly aggregated and shown on a screen with no direct relation to the physical instrument.
We investigated different modalities and visual encodings to represent feedback for songs or exercises directly on the instrument in more detail and to allow for visual comparison between exercise repetitions.
While augmented reality (AR) has been used for instrument feedback before~\cite{yamabe2011feedback, loechtefeld2011guitar, skreinig2022arhero, gutierrez2020augmented, huang2011piano, takegawa2008piano, das2017music, johnson2019evaluating}, the visual encodings are simple and focus only on immediate feedback.
Our AR visualization designs comprise larger amounts of data. In terms of instruments, we target drum kits and guitars.

\section{Related Work}

Research on music-related visualization~\cite{khulusi2020survey} explored encodings for structure~\cite{chan2007report, watanabe2003brass}, listening histories~\cite{baur2010streams}, collections~\cite{miller2022corpusvis}, and theory~\cite{
miller2022augmenting, bunks2022jazz}. 
As practice data has gotten little attention yet~\cite{heyen2022instrudata}, we focus on amateur exercises and visual comparison
between personal
recordings. 
Work in human-computer interaction explored supporting instrument learning through assessments
that can be visualized within sheet music~\cite{asahi2018toward:piano:support, hori2019piano:hmm} or on instruments with LEDs~\cite{marky2021letsfrets} or projector-based augmented reality displays~\cite{yamabe2011feedback, loechtefeld2011guitar}.
Similar to apps and games\footref{apps}, these approaches only provide simple immediate feedback, as they do not involve detailed visualization.
Virtual and augmented reality (VR/AR) has been used to teach dancing~\cite{chan2010virtual}, guitar~\cite{motokawa2006support, skreinig2022arhero, gutierrez2020augmented}, piano~\cite{huang2011piano, takegawa2008piano, das2017music}, theremin~\cite{johnson2019evaluating}, and drums~\cite{yamabe2011feedback} and to visualize movements of cellists~\cite{heyen2022cellovis}.
We extend this work with further visualizations that show more than what is currently played by leveraging immersive analytics, which combines VR/AR with visualization,
and situated analytics,
where information is displayed near a physical object it relates to.

\section{Design}

We conducted two design studies that explore AR visualization for real-time and post hoc feedback, on either drums or guitar.
For drums, we used projector AR and glanceable
visualizations, for guitar an optical see-through HMD with 3D visualizations on or next to the instrument.

\subsection{Projector-Based AR for Drums}

\begin{figure}[h]
  \centering
  \includegraphics[width=0.75\linewidth]{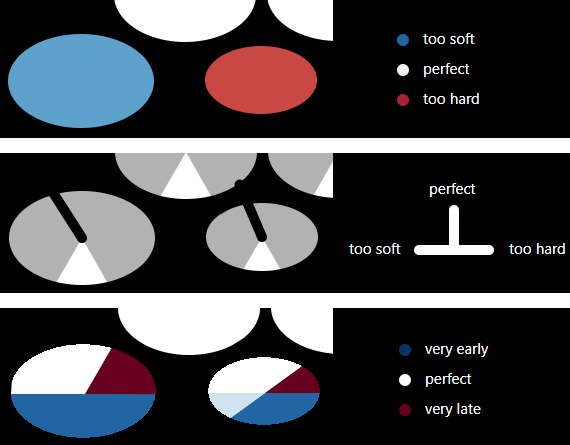}
  \caption{
  Comparison of drum visualizations.
  }
  \label{fig:drum_vis}
\end{figure}

Our hardware setup consist of a small projector (Elephas GC333) mounted onto a tripod, with a camera for easier calibration (\autoref{fig:teaser}).
We use a mobile drum kit (Millenium MD-90), as its pads are at similar distance to the projector and therefore are simultaneously in focus.
To improve the visibility of projections, we attached white paper to the pads.
The drum kit outputs MIDI data, from which we use timing and velocity (strength of hits), to tell users whether they were too early/late or too soft/hard, depending on the chosen mode.
We designed three visualizations to reveal this information quickly (\autoref{fig:drum_vis}):
The first one fills the drum's pad with a color after each hit, either until the next hit or only for a short flash.
As color perception is limited, we added a tachometer-inspired encoding with a needle that goes straight up for correct hits, towards the left for too soft/early and towards the right for too hard/late.
The third visualization summarizes previous playing through pie charts, showing the ratio of, for example, different classes of timing deviations.

\subsection{Optical See-Through HMD AR for Guitar}

\begin{figure}[h]
  \centering
  \includegraphics[width=0.6\linewidth]{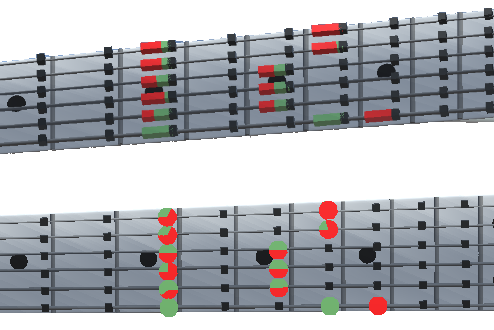}
  \caption{
  Visualizations can be flat and simple, as these bar and pie charts, or 3D like the stacked bars in \autoref{fig:teaser}.
  }
  \label{fig:bars_and_pies}
\end{figure}

\begin{figure}[h]
  \centering
  \includegraphics[width=0.82\linewidth]{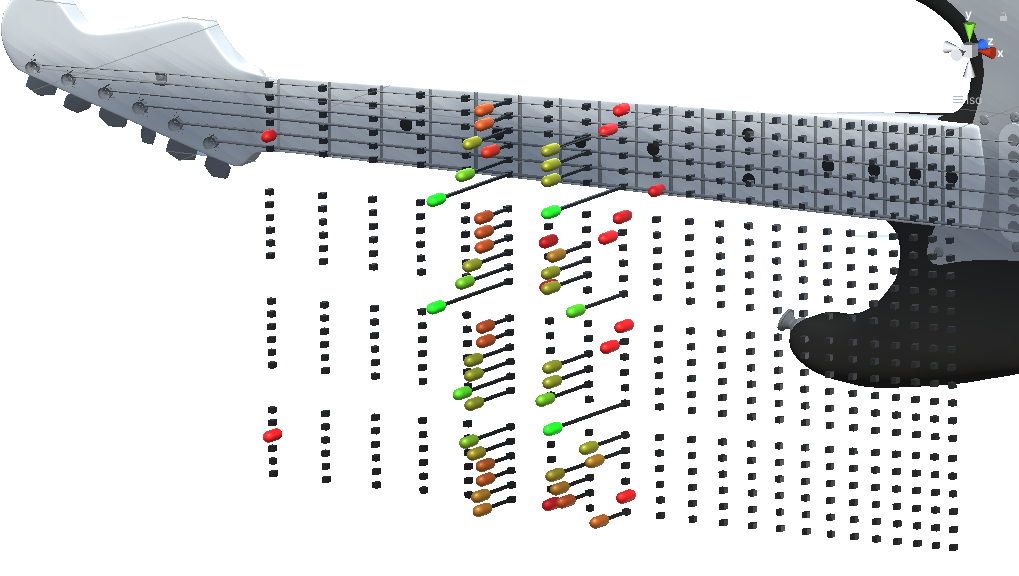}
  \caption{
  As we are using a screen as mirror or an HMD, we can display information not only on, but also around the instrument.
  In this example, juxtaposed multiple views of different exercise recordings allow comparison.
  }
  \label{fig:juxtaposed_3d_vis}
\end{figure}

As projectors are limited to displaying two-dimensional information~\cite{loechtefeld2011guitar}, we chose a head-mounted display (Microsoft HoloLens) for our next design study.
We fitted a guitar with printed optical markers to be able to track it.
As an alternative modality, we also implemented mirror-AR, using a PC's screen to display visualizations as overlay on a mirrored webcam image.
Our example data includes scale exercises along with how often notes were played in total, how many were missed, and how many were played extra.
We can further show timing errors, for example binned into different error severity classes.
The visual encodings include flat pie and bar charts (\autoref{fig:bars_and_pies}), 3D bar charts, and ''balloons`` (\autoref{fig:juxtaposed_3d_vis}) that reduce occlusion but still retain relation to fretboard positions.
We implemented different ways to arrange the virtual fretboard and its attached visualization:
As the guitar's fretboard is hardly visible when holding the guitar in playing position, users can move the virtual fretboard in any direction, or flip it upwards.
Multiple fretboards can be shown stacked next to each other, for better comparison (\autoref{fig:juxtaposed_3d_vis}).

\section{Discussion}

Our AR visualizations show data, right where it belongs to, with more details then in related work and in the case of our guitar visualizations also around the instrument.
We hypothesize that this could make information easier to understand, as less context switches are needed. Still, this approach also bears some limitations.
We encountered several technical challenges, such as the projector's low resolution and focus range, which limit the complexity of visualizations and lead to blur.
To address this limitation, we designed simplified encodings and use a mobile drum kit where pads are in the same plane.
Our AR HMD has a small field of view, so visualizations were not entirely visible at once when the guitar was held in playing position.
Our mirror screen did not have this problem, but suffered from the same tracking issues, where visualizations and guitar where sometimes misaligned.
A general drawback of live visualizations is that they distract from playing, even if they are glanceable.
Three-dimensional visualizations can lead to occlusion, although this has less impact when using a stereoscopic HMD. 
As this work is preliminary, we did not evaluate our design and implementation with other potential users.
Ideally, a refined version should be tested with musicians over several weeks in their daily or weekly practice.

\section{Acknowledgments}

This work was funded by the Cyber Valley Research Fund.

\pagebreak

\bibliography{ISMIRtemplate}

\end{document}